\documentclass{elsart}
\usepackage{natbib,graphicx}
\begin{document}
\begin{frontmatter}
\hfill WUB-99-16\\
\title{Stochastic Estimator Techniques and Their Implementation on
Distributed Parallel Computers}
\author{Stephan G\"usken}

\address{Bergische Universit\"at Wuppertal, Fachbereich Physik,
42097 Wuppertal, Germany}

\begin{abstract}
The calculation of physical quantities by lattice
QCD simulations requires in some important cases the determination
of the inverse of a very large matrix. 
In this article we describe how stochastic estimator methods can be
applied to this problem, and how such techniques can be efficiently
implemented on parallel computers. 
\end{abstract}
\end{frontmatter}

\section{Introduction}

Within our current level of comprehension of the fundamental principles
of nature, physical  processes on an atomic or subatomic scale can be
successfully described by Quantum Field Theories (QFT). In such theories
particles as well as their interactions are represented by quantum
fields, defined at each space-time point. The value of a physical 
quantity, which
can be measured in experiment, can be calculated by a weighted
average over all ``would be'' values of this quantity, achieved for 
each possible configuration of the quantum fields involved.
The weight with which each of these ``would be'' values contributes
is determined by the so called action, a scalar quantity which 
contains the characteristic features of the QFT in question 
and which depends on the quantum fields. The formal expression
of this averaging procedure is known as the path functional.   
  
An exact analytical treatment of the path functional is in most
cases not possible. Approximate solutions can be achieved 
in the framework of perturbation theory if the interaction strength
is weak.
Perturbative methods have been proven very successful in the evaluation of
the QFT of the electromagnetic and weak forces. They fail however
when applied to the QFT of the strong force, the so called Quantum
Chromo Dynamics (QCD). The strong force is responsible for a large
range of phenomena at and below  the scale of the atomic nuclei.

Lattice QCD is designed for a non-perturbative numerical
evaluation of QCD. The space-time continuum is approximated by 
a lattice with $N_s^3 \times N_t$ space-time points. 
The calculation of physical quantities is done in two steps.
First, one generates a representative sample of quantum field
configurations, where each configuration is represented according to
its specific weight, by a Monte Carlo
procedure \cite{mc_quenched,mc_full,Kennedy_here}. Secondly, one determines 
the ``would be'' value of the physical quantity in question
on each of the quantum
field configurations and takes the average. We call the latter step the
analysis of quantum field configurations.

Clearly, the computational effort which has to be invested in the
analysis part depends on the physical quantity one is
interested in. 
In this article we report on a computationally very hard problem
which occurs in the analysis of configurations with respect to
so called disconnected contributions. The latter are of great physical
importance as they are expected to play  a substantial role in the
solution of the ``proton spin problem'' \cite{review_ga} and of the
``$U(1)$ problem'' of QCD \cite{reviews_u1}. 

Naively, the
calculation of disconnected contributions  requires the inversion
of a complex matrix of size 
$(N_s^3 \times N_t \times 12)^2$ on
each single quantum field configuration.
For currently available lattice sizes, see ref.\cite{burkhalter}, such
a calculation would be prohibitively expensive.
To circumvent this problem one applies stochastic estimator methods which 
converge to the true result in the stochastic limit.
It turns out however, that even with such techniques one still needs
a parallel supercomputer to handle this problem.

This article is organized as follows.
In the next section we will give an impression of the physical
meaning of disconnected contributions.
Section 3 explains what has to be done technically
to calculate such contributions.
Section 4 is devoted to the stochastic estimator techniques.
The implementation of these techniques on parallel computers
is discussed in section 5.
Section 6  will give an overview of state of the art calculations of
disconnected contributions. Section 7 contains a short summary.
 
\section{The Physical Motivation}

It is nicely explained by R. Gupta in this volume \cite{gupta_here}
that our ``parton'' picture of a proton as being made of 3 interacting
quarks is not applicable in full QCD. The reason is that in this case
the spontaneous creation and annihilation of quark antiquark pairs
leads to an additional contribution to the proton amplitude.
This is shown in Gupta's fig.3.

Suppose that we would like to investigate the properties of
a proton by a scattering experiment, 
e.g. by deep inelastic $\mu$p (muon proton) scattering. Then, 
the $\mu$p scattering
amplitudes measured in such an experiment could in principle
differ sizably from the parton expectation since the latter neglects the
interaction of the $\mu$ particle with the quark antiquark loop.  

\begin{figure}
\begin{center}
\includegraphics*[scale=1.0]{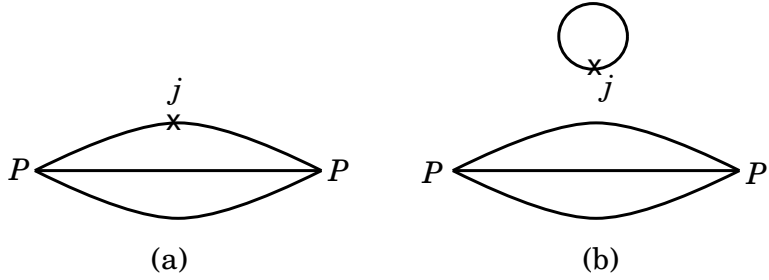}
\end{center}   
\caption{\label{fig_schema_con_dis}{\it Connected (a) and disconnected (b)
contributions to a proton interacting with a current $j$. 
All quark lines, including the quark loop, are connected by infinitely
many gluon lines and virtual quark loops. 
}} 
\end{figure}

To illustrate this point we show in fig. \ref{fig_schema_con_dis} the
full QCD contributions to the propagator of a proton which interacts
with an external current $j$.
Part (a) of the figure depicts the naive (parton) case, where the current
couples to one of the quarks of the proton. Part (b) shows the
interaction of the current $j$ with a quark antiquark loop, in the
field of the proton.
This disconnected contribution is present only in full
QCD. We emphasize  that the location in space and time of
the quark antiquark loop is
not fixed. Thus, to calculate the disconnected contribution one has to
sum over all positions.

Of course, it is not obvious from these considerations that the
disconnected part really gives non negligible contributions to the
scattering amplitudes. In fact, it turns out that many of them
have a structure such that their net contribution is expected
to be small.

There is however a class of amplitudes, the flavor singlet
amplitudes, where disconnected contributions can be sizeable. In order to 
investigate quark loop effects in QCD it is therefore of utmost
interest to calculate flavor singlet amplitudes and to compare
the results with experimentally measured data. 

From experimental measurements one can extract the values of at least
2 flavor singlet amplitudes. The first, which describes the
interaction of a proton with a pseudo-vector current deviates
by about a factor of 2
from the naively expected value. This deviation gave rise to the so called
``proton spin crisis''. The second, which couples a scalar current to
a proton, yields, when multiplied by the quark mass, the pion-nucleon
sigma term $\Sigma_{\pi N}$ \cite{gasser_nsigma}. The experimental
value of this quantity also differs by about a factor of 2 from the
naive expectation. Thus, these quantities are most promising
candidates to study the influence of quark loops by a full QCD lattice
simulation.

\begin{figure}
\begin{center}
\includegraphics*[scale=1.0]{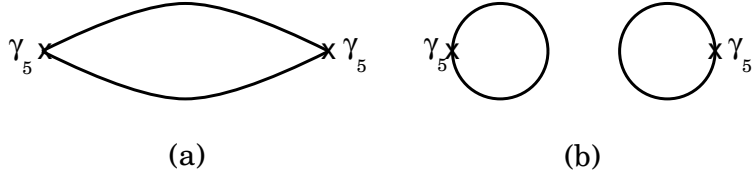}
\end{center}   
\caption{\label{fig_schema_eta}{\it Connected (a) and disconnected (b)
contributions to the propagator of a pseudo scalar meson.}} 
\end{figure}

Disconnected amplitudes are supposed to contribute also to many
physical processes other than proton scattering. For example,
a (pseudo scalar) meson, which is made of a quark and an antiquark,
can be ``mimicked'', with respect to its quantum numbers, by 2
quark antiquark loops. This is shown in fig.\ref{fig_schema_eta}.  
An experimental measurement of e.g. the mass of this meson
would include both terms, connected and disconnected.

From symmetry
considerations one again concludes, that the disconnected
part should contribute mostly if the quarks in the meson
are put together in a flavor singlet combination.   
Experimentally, one finds that the mass of such a flavor singlet
meson, which is named $\eta'$, is much larger than that of its
non-singlet partners. This discrepancy is called the ``$U(1)$ problem 
of QCD''.

Clearly, a full QCD lattice calculation of the diagrams of
fig. \ref{fig_schema_eta} would be of great help to solve this problem.

\section{The Technical Problem}

\subsection{Quark Propagator}

A key quantity in the analysis of quantum field
configurations is the quark propagator $\Delta$.
It is defined as the correlation of 2 (fermionic) quantum fields 
$\Psi$ and $\bar{\Psi}$ at the space-time points $(\vec{x},t)$ and
$(\vec{x}',t')$ respectively:
\begin{equation}
\Delta(\vec{x},t,a,\alpha;\vec{x}',t',a',\alpha') =
 \psi(\vec{x},t,a,\alpha)\bar{\psi}(\vec{x}',t',a',\alpha') \;.
\label{eq_qprop_def}
\end{equation}
The indices $a,\alpha$, $a',\alpha'$ denote internal degrees of
freedom of the fermionic quantum fields. $a$,$a'$ are called color
indices. They can take the values 1,2 and 3. $\alpha$,$\alpha'$ are
called Dirac indices. They run from 1 to 4.

In the language of QCD, the quark propagator $\Delta$ denotes
the probability amplitude of a strongly interacting elementary
particle (quark) to travel from point $(\vec{x},t)$ to point
$(\vec{x}',t')$. Once $\Delta$ is known, a whole bunch a physical
quantities like the spectrum and the decay properties of strongly
interacting composite particles can be determined immediately.  
  
Unfortunately, eq. (\ref{eq_qprop_def}) cannot be used for a numerical
calculation of $\Delta$ since, with current Monte Carlo algorithms,
the fermion fields $\Psi,\bar{\Psi}$ are not explicitely available.
They enter only indirectly in form of the fermionic matrix $M$. The
quark propagator is related to $M$ by 
\begin{equation}
\Delta(x;x') = M^{-1}(x,x') \;,
\label{eq_qprop_M}
\end{equation}
where we have used the multi index $x$ for space-time, color and Dirac
indices. 

The fermionic matrix $M$ is complex and sparse. In the widely used
Wilson form it is given by
\begin{eqnarray}
\lefteqn{M(x,a,\alpha;x',a',\alpha') =} \label{eq_M_Wilson} \\
&& \qquad \qquad 1 - \kappa \sum_{\mu=1}^4 
\left[
(1 - \gamma_{\mu}(\alpha;\alpha'))U_{\mu}(a;a')(x)
\delta_{x+\hat{\mu};x'} + \right. \nonumber \\
&& \qquad \qquad \qquad \qquad \;
(1 + \gamma_{\mu}(\alpha;\alpha'))U_{\mu}^{\dagger}(a;a')(x-\hat{\mu})
\delta_{x-\hat{\mu};x'} \left. \right] \;. \nonumber
\end{eqnarray}
Here, $\kappa$ is a real number, which determines the mass of the
propagating quark. $\gamma_{\mu}$ denotes the $4 \times 4$ 
anti-commuting Dirac matrices. The ``links'' $U_{\mu}$ are $SU(3)$ matrices,
which act in color space. They represent the quantum fields of the
interaction between quarks. The unit vector $\hat{\mu}$ points into
the direction $\mu$.

According to eq. (\ref{eq_qprop_M}), the computationally expensive
part of the analysis is to determine the inverse of $M$ for each
quantum field configuration $U$. Fortunately, for many applications, 
it is not necessary to solve the full problem. For example, to
calculate the spectrum and the decay properties of strongly
interacting  composite particles, it is sufficient to determine only one
row of $M^{-1}$. This reduced problem
\begin{equation}
M(z,x) \Delta(x,x_0) = \delta_{z,x_0} \quad,\quad x_0 \; \mbox{fixed}
\label{eq_1rowprop}
\end{equation}  
can be treated using fast iterative solvers \cite{Thomas_here} with
moderate computational effort.
 
\subsection{Disconnected Contributions}

There is however a  class of important physical quantities (see
above), whose determination requires, in a sense, the solution of
the full problem. To be specific, the prominent combinations
$D_{\Gamma}$ of
$M^{-1}$ needed for the calculation of the
disconnected contributions are given by
\begin{equation}
D_{\Gamma} = Tr \left[ \Gamma M^{-1} \right] \;,
\label{eq_def_DG}
\end{equation}
where $\Gamma=1,\gamma_{\mu},\gamma_5,\gamma_{\mu}\gamma_{5}$ 
is a $4\times 4$ matrix which acts on the Dirac indices of $M^{-1}$.
$\gamma_5$ is defined by $\gamma_5 = \gamma_1\gamma_2\gamma_3\gamma_4$.
  
Clearly, an exact determination of $D_{\Gamma}$ would require 
$N_s^3 \times N_t \times 3 \times 4$ applications of the ``row'' 
method, eq. (\ref{eq_1rowprop}). This would  overtax even the capacity
of a fast parallel supercomputer.

We will see in the next section how one can circumvent this problem
by a calculation of a reliable estimate, $D^E_{\Gamma}$, instead of the
exact value $D_{\Gamma}$.

\section{Stochastic Estimation}

Suppose that we would have created $N$ random vectors 
$\eta_k(i)$,  $i=1,\dots,V$, $k=1,\dots,N$  with the properties
\begin{eqnarray}
\lim_{N \rightarrow \infty} \frac{1}{N} \sum_{k=1}^{N} \eta_k(i) 
&\equiv & \langle \eta(i)\rangle =0 \;, \label{eq_stochdef_1} \\
\lim_{N \rightarrow \infty}\frac{1}{N} \sum_{k=1}^{N} \eta_k(i)\eta_k(j)
&\equiv & \langle \eta(i) \eta(j)\rangle = \delta_{i,j} \;. 
\label{eq_stochdef_2}
\end{eqnarray}
These properties are fulfilled by, for example, Gaussian 
\cite{bitar_gauss} 
or $Z_2$ \cite{liu_z2,liu_z2_sup} random number distributions.

Suppose furthermore that we would modify eq. (\ref{eq_1rowprop}) by
inserting a random
source vector $\eta_k$ on the right hand side
\begin{equation}
M(z,x) \tilde{\Delta}_k(x) = \eta_k(z) \;,\; 
\tilde{\Delta}_k(x)=\delta_{x,x'}\eta_k(x') \;.
\label{eq_etaprop}
\end{equation}  
Then, the product $\eta_k \tilde{\Delta}$ can be written as
\begin{equation}
\tilde{D}_1^k = Tr(\Delta) \sum_{i=1}^{V} \eta(i)_k\eta(i)_k +
\sum_{i \neq j} \Delta(i,j)\eta(i)_k\eta(j)_k \;.  
\label{eq_stochest_dk}
\end{equation}
According to eq. (\ref{eq_stochdef_2}) one gets in the stochastic
limit ($N \rightarrow \infty$) of $N$ solutions to
eq. (\ref{eq_etaprop})
\begin{equation}
\langle \tilde{D_1}\rangle = D_1 \;.
\end{equation}
Thus, this procedure converges to the correct result. We mention that
the stochastic estimator method can be also applied, with small
modifications \cite{sesam_nsigma}, to the calculation of arbitrary 
$D_{\Gamma}$, c.f. eq. (\ref{eq_def_DG}).
 
Of course, the stochastic method is useful only if
already a moderate number $N$ of solutions to  eq. (\ref{eq_etaprop})
suffices to calculate a reliable estimate
\begin{equation}
D_1^E = \frac{1}{N}\sum_{k=1}^N \tilde{D}_1^k
\label{eq_D_estsum} 
\end{equation}
of $D_1$. The question of how large $N$ should be chosen has
been investigated by the authors of  ref. \cite{sesam_disc} in
some detail for a medium size lattice ($V=16^3\times 32 \times 3
\times 4$). It turned out that $N \simeq 100$ allows to estimates  
$D_1$ within a $10\%$ uncertainty. The situation might be 
much less favorable however  for $\Gamma \neq 1$. For example, for
\begin{eqnarray}
\lefteqn{ D_{\Gamma} = D_{\gamma_3\gamma_5}  = } \\ 
&& \qquad \qquad \sum_{x,a} \left[\; \Delta(x,a,1;x,a,1) - \Delta(x,a,2;x,a,2)
\right. \nonumber \\
&&  \quad \qquad \qquad \, - \Delta(x,a,3;x,a,3) + \Delta(x,a,4;x,a,4)
 \left. \right] \nonumber
\end{eqnarray}  
one has to determine differences of the diagonals of $\Delta$ instead
of the sum over diagonal elements. Since all these numbers are of
similar size, such a task could require a much higher number of
estimates to achieve a reliable result on each single quantum field
configuration.

Fortunately, the problem is softened by the average over quantum field
configurations, for the following reason: Quantum Field Theories, such
as QCD, exhibit the property of gauge invariance, i.e. physical
quantities do not change their values  under gauge transformations. The path
integral, which represents the formal expression of how to calculate
physical quantities in QFT, automatically removes all non gauge
invariant contributions. Since most of the (unwanted) noise terms 
on the right hand side of eq. (\ref{eq_stochest_dk}) are not gauge
invariant, the average over quantum field configurations will help
to increase the accuracy of the estimate of
$D^E_{\Gamma}$. Nevertheless, as we will show at the end of this
article, one still needs at least 400 estimates per quantum field
configuration to achieve statistically significant signals for
$D_{\Gamma}$ or for the (physically important) correlations between 
$D_{\Gamma}$ and the proton propagator.

Thus, the computational effort which is necessary to calculate
disconnected contributions exceeds the one for the ``standard
analysis'' by more than 2 orders of magnitude. 

\section{Parallelization}

There are a least two straightforward ways to implement the numerical
problem defined by eqs. (\ref{eq_etaprop}),(\ref{eq_D_estsum})
on a parallel computer. The first, which we call 
``external parallelization'' can be used for medium size lattices on
machines with a, compared to the processor speed, slow communication
network. The second one, which we
name ``internal parallelization'' is useful for large lattice and, on
machines with fast communication lines.

\subsection{External Parallelization}

A natural way to implement the stochastic estimator method on a
parallel computer arises from the fact that the estimates, 
eq. (\ref{eq_etaprop}), are completely independent of each other. 
Thus, the estimates can be calculated simultaneously on separate
compute nodes. Communication is required only at the beginning of the
calculation, when quantum fields and stochastic sources have to be
passed to their respective nodes, and at the end, when the results of
the single estimates have to be gathered and averaged.  

Alternatively, one can implement ``external parallelization'' with
respect to the quantum field configurations. In this case, each
processor receives its own quantum field configuration at the
beginning and computes $N$ estimates. 

In both cases, one has to ensure that the random numbers used on such
a distributed system are not correlated. This can be achieved either
by running a large period random number generator only on one node,
which passes the random vectors successively to all other nodes, or
by using a parallel random number generator, where each node creates
its own random numbers from an independent stream of the generator 
\cite{random_number_generators}.   
   
An ideal machine to implement on a stochastic estimator program in
the ``external'' mode would be a large cluster of powerful
workstations, which are connected e.g. by Ethernet. 

Let us give an example. One needs with a standard inverter, say the
minimal residual (MR) inverter, on a workstation which runs with a
sustained speed of 50 Mflops about 20 minutes of CPU time to solve
eq. (\ref{eq_etaprop}) on a $16^3 \times 32$ lattice for values 
of $\kappa$, c.f. eq. (\ref{eq_M_Wilson}), in the physically
interesting range. Thus, on a cluster of 100 workstations it would take 
about 11 days to calculate $D^E_\Gamma$ with 400 estimates on 200
quantum field configurations. 

The memory requirements on each node for medium size lattices are moderate.
For a $16^3 \times 32$ lattice one needs an overall amount of
about 60 Mbytes. Thus, even a $26^3 \times 52$ lattice would easily
fit into the memory of a 512 Mbyte workstation. 
 
\subsection{Internal Parallelization}
 
\begin{figure}
\begin{center}
\includegraphics*[scale=0.7]{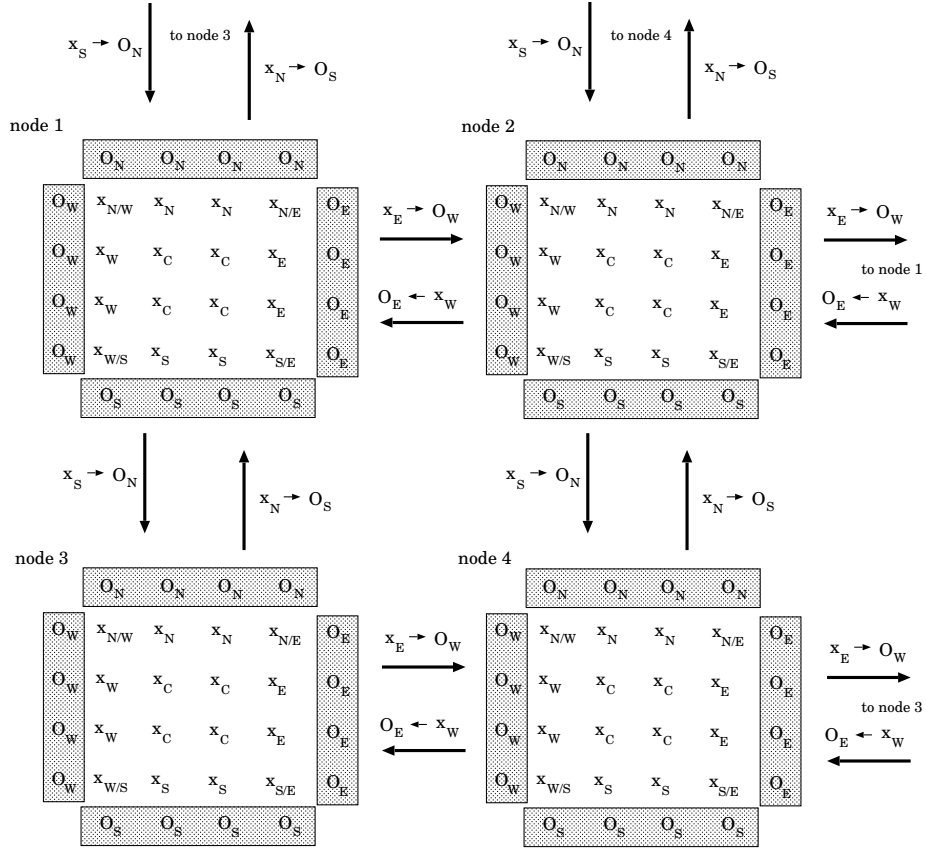}
\end{center}   
\caption{\label{fig_scheme_para}{\it Internal parallelization of a
$8 \times 8$ lattice with periodic boundary conditions lattice on 
4 node system.
 }} 
\end{figure}

To handle large lattices on parallel computers with  a comparatively
small amount of memory per node or on massively parallel systems with
a large number of nodes, one should divide the lattice into
sub-lattices and distribute the latter among the nodes. Since 
the matrix $M$, see eq. (\ref{eq_M_Wilson}), 
connects only nearest neighbors, this can be accomplished in a
straightforward way. A simple but efficient realization of this
``internal parallelization'' is shown in fig. \ref{fig_scheme_para}.
for an $8 \times 8$ lattice. Each node administers the data points of 
a $4 \times 4$ sub-lattice (denoted by crosses) as well as the
current values of the surface points of the neighboring sub-lattices
(denoted by O). After each iteration step of the solver, e.g. MR, 
the updated values of each sub-lattice ($X_N,X_S,X_W,X_E$) are passed
to the ``O-buffers'' of the neighboring sub-lattices, i.e.
$X_S \rightarrow O_N, X_N \rightarrow O_S$ etc. . This procedure is
repeated until some stopping criterion, set by e.g. an upper bound
of the norm of the rest vector, is fulfilled.  
  
The ``internal parallelization'' requires of course a communication
network of much higher quality than the ``external
parallelization''. Although the amount of data which has to be exchanged
between the nodes after each iteration step can be adjusted by a
suitable choice of the sub-lattice size, the number of communications
necessary to complete one full estimate can not: It is determined by
the number of iterations needed by the solver to converge. Typically,
this number is in the range of a few hundred for $\kappa$ values in
a physically interesting region. Thus, the startup time for the
communication has to be taken seriously  into account.
 
We mention that the memory overhead introduced by the ``O-buffers''
can be avoided on shared memory machines. On such a machine each processor
reads the required data directly from the memory of the neighboring
nodes.

Finally, we emphasize that ``external'' and ``internal''
parallelization can be mixed. In this case one would divide the total of
nodes into subgroups, on which one would implement ``internal''
parallelization. These subgroups could then run essentially
independent of each other, in the ``external'' mode,
to work at different estimates.
An advantage of such a mixed solution is that one needs fast
communication only within the subgroups. Furthermore, one would
achieve more flexibility in optimizing the ratio of communication
to CPU time on a given set of nodes.
 
\section{What has been achieved so far}

The evaluation of disconnected contributions with stochastic estimator
methods is still at its beginning. Clearly, this is due to the 
requirement of a very high computer speed needed to tackle such a
problem. 

Pioneering studies of disconnected contributions have been
performed some time ago on conventional (vector) computers in the
quenched approximation of QCD, where one neglects the fermionic
quantum fields in the Monte Carlo update, by the authors of refs.
\cite{japan_nsigma,japan_ga} and \cite{liu_nsigma,liu_ga}. 
Due to the lack of computer speed, the authors had to make concessions
to the statistical reliability of their results. Thus, although
the data look promising, a systematic bias of their findings cannot be
excluded.

With the advent of powerful parallel computers it became possible to  
treat disconnected contributions more reliably  
 \cite{sesam_nsigma,sesam_disc,sesam_ga,kilcup}, although one is still 
limited to medium size lattices.

So far, the most intense study of such contributions has been
performed by the SESAM collaboration \cite{sesam_nsigma,sesam_ga} on 
a QH2 APE-100 computer \cite{rapuano_here} which runs the
stochastic estimator code  with a sustained
speed of $\simeq 6$ Gflops. SESAM has analyzed 200 full QCD quantum field
configurations of a $16^3 \times 32$ lattice, at several values of
the mass parameter $\kappa$. On each configuration, and for each $\kappa$, 
the values of $D_1$ and $D_{\gamma_3\gamma_5}$ have been estimated 
400 times.
   
The statistical analysis of the SESAM data revealed that the
related physical quantities, i.e. the correlation $C_{\Gamma}$ 
of $D_{\Gamma}$ and
the proton propagator with respect to the quantum field
configurations, can be determined within an uncertainty of 
$30\%$ for $C_1$ and $50\%$ for $C_{\gamma_3\gamma_5}$ within this 
setup.

Clearly, this is not satisfactory. But, as we pointed out above, the
use of stochastic estimator techniques is still in its infancy. 
The SESAM result sets the stage of what has to be invested 
to achieve the goal of calculating disconnected contributions
within a few percent uncertainty.

Besides the expected increase of computational power over the next
years, improvements of the stochastic estimator technique will help to
increase the statistical significance in the calculation of
disconnected amplitudes. Promising suggestions along this line can be
found in \cite{philippe} and \cite{viehoff_dolby}.

\section{Summary}

We have illustrated that the calculation of disconnected contributions with
stochastic estimator methods represents a computationally very hard 
problem in lattice QCD.

Fortunately, there are several straightforward ways to implement the
code on a parallel computer. Thus, almost every powerful parallel machine can
be used.

In view of the intrinsic parallelism of the problem with respect to
the number of estimates, the ideal computer to run the code for medium
size lattices is a large cluster of workstations.

\end{document}